\documentstyle[twoside,12pt,a4]{article}
\newcommand{\btab}{\begin{tabbing}}
\newcommand{\etab}{\end{tabbing}}

\newcommand{\beqn}{\begin{equation}}
\newcommand{\eeqn}{\end{equation}}
\newcommand{\barr}[1]{\begin{array}{#1}}
\newcommand{\earr}{\end{array}}
\newcommand{\beqna}{\begin{eqnarray}}
\newcommand{\eeqna}{\end{eqnarray}}
\newcommand{\btablec}{\begin{table} \begin{center}}
\newcommand{\etablec}{\end{center} \end{table}}
\newcommand{\lapprox}{\stackrel{<}{\scriptstyle \sim}}
\newcommand{\gapprox}{\stackrel{>}{\scriptstyle \sim}}
\newcommand{\gapproxeq}{\lower.7ex\hbox{$\;\stackrel{\textstyle>}{\sim}\;$}}
\newcommand{\lapproxeq}{\lower.7ex\hbox{$\;\stackrel{\textstyle<}{\sim}\;$}} 
\newtheorem{theorem}{Theorem}
\newcommand{\bth}{\begin{theorem}}
\newcommand{\eth}{\end{theorem}}

\newcommand{\plabel}[1]{\label{#1}}
\newcommand{\pbibitem}[1]{\bibitem{#1}}
\newcommand{\jpsi}{{J/\psi}}
\input epsf

\begin{document}
\title{\begin{flushright}  
\small{hep-ph/9610250}\\
 \small{MC-TH-96/28} \\ \small{RAL-96-075}
 \end{flushright} 
\vspace{0.6cm}  
\LARGE \bf Gluonic Mesons in $\jpsi$ Radiative Decay}
\author{Philip R. Page\thanks{E--mail: prp@a13.ph.man.ac.uk.}\\
{\small \em Department of Physics and Astronomy, University of Manchester,} \\
{\small \em Manchester M13 9PL, UK} \\
Xue--Qian Li\thanks{On leave from the Department of Physics, Nankai University, Tianjin
300071, P.R. China.}\\
{\small \em Particle Theory, Rutherford Appleton Laboratory, Chilton,
Didcot OX11 0QX, UK}}
\date{October 1996}
\maketitle
\abstract{The $\eta(1760)$ may be significantly produced in $\jpsi$
radiative decay. We deduce from experimental data that its branching ratio 
to two gluons is bounded,
$0.4\pm 0.3 \lapprox BR(\eta(1760)\rightarrow gg)\lapprox 0.9\pm 0.3$, 
consistent with gluonic admixture. 
Moreover, the expectation that there should be an isoscalar analogue
of the gluonic (hybrid) meson candidate $\pi(1800)$ is argued to reinforce 
hybrid admixture in $\eta(1760)$.  
Two--gluon coupling of hybrids is estimated in a spectator model
to be considerably larger than that of $Q\bar{Q}$. In addition,
the two--photon coupling of $\eta(1760)$ is estimated to be 
$1.4\pm 0.7$ keV.}

\newpage

We cannot be content whilst the strong dynamics of gluonic degrees of freedom
remain an area of substantial ignorance. The existence of glueballs and hybrids 
(mesons with explicit gluonic excitation) remain unconfirmed.  
$\jpsi$ radiative decay has yielded glueball candidates
$f_0 (1500), \; f_J (1710), \; f_J (2220)$ \cite{bugg95} and would {\it a priori} serve as an
ideal isoscalar gluonic meson production mechanism.
It is tempting to restrict analysis of the branching ratio of a resonance $R$ into two gluons
to the expected extremes of the allowed range, i.e. to glueballs
where $BR (R \rightarrow gg)\sim 0.5 - 1$ or to quarkonia  
where $BR (R \rightarrow gg)\sim O(\alpha_S^2) \approx 0.1-0.2$  \cite{farrar96}. 
It is, however, imperative to consider the possibility of hybrid admixture. 
This is especially true in the pseudoscalar sector  where both states in 
$\eta(1440)$ possess $BR(\eta(1440)\rightarrow gg) = 0.5 -1$ \cite{farrar96,cakir94}, not
compatible with pure $Q\bar{Q}$. We are thus motivated to determine the two--gluon 
coupling of hybrids in a model.

In this letter we shall focus on $\eta(1760)$. Firstly, we investigate its radiative
production in $\jpsi$ decay, and show how this can be used to 
obtain a lower bound on the two--gluon coupling of
$\eta(1760)$. The bound is consistent with a gluonic admixture in the state. 
We then show that a sensible and constrictive upper bound on the two--gluon coupling of  $\eta(1760)$
can be obtained from $\Upsilon$ radiative decay, corroborating the validity
of the methods used. The expectation that there should be an isoscalar analogue
of the hybrid candidate $\pi(1800)$ is argued to reinforce 
gluonic admixture in $\eta(1760)$. Moreover, we then expect
hybrid as opposed to glueball admixture. We proceed to make theoretical estimates
for the two--gluon coupling of hybrids and $Q\bar{Q}$, and indicate that hybrid
coupling is substantially larger than that of $Q\bar{Q}$, consistent with expectations.
Lastly, we show how the two--photon coupling of $\eta(1760)$ can be used in conjunction with its two--gluon 
coupling to establish whether $\eta(1760)$ is dominantly gluonic or $Q\bar{Q}$. The two--photon coupling is
estimated using vector meson dominance, showing that detection is a realistic prospect.

Experimentally, $\eta(1760)$ has a
mass of $1760\pm 11$ MeV and a width of $60\pm 16$ MeV, as
well as $BR(\jpsi\rightarrow\gamma \eta(1760))\; BR(\eta(1760) \rightarrow
\rho^0\rho^0) = (0.13\pm 0.09) \times 10^{-3}$ \cite{pdg96}, all derived from 
DM2 data \cite{rho}. Even though these results are used  throughout,
they are in need of confirmation. Mark III data \cite{rho} on
$\eta(1760)\rightarrow\rho\rho$ have been re--analysed \cite{bugg95}
incorporating the neglected $(\pi\pi)_S\; (\pi\pi)_S$ mode in the original analysis. The re--analysis
eliminated the $J^{PC}=0^{-+}$ resonance, indicating that improved data analysis
of DM2 data is also needed, although it was admitted that the parametrization used 
in the re--analysis ``may be patching up $0^{-}$ contributions well above 1500 MeV'' \cite{bugg95}. 
It is, however, significant that
$\eta(1760)$ was initially observed in the relatively clean $\omega\omega$
channel \cite{omega}. 

A discriminator between the gluonic and $Q\bar{Q}$ nature of  $\eta(1760)$ is provided
by its two--gluon coupling. In refs. \cite{farrar96,cakir94} a formalism was presented
which successfully connects the two-gluon 
width $\Gamma (R \rightarrow gg)$ of a pseudoscalar resonance to the radiative $\jpsi$
branching ratio $BR(\jpsi\rightarrow\gamma R)$. Here 

\beqn \plabel{ps} 10^3 BR(\jpsi\rightarrow\gamma R) = \frac{M_{R}}{1.8\;
GeV}\frac{\Gamma (R \rightarrow gg)}{50\;
MeV}\frac{x|H_{PS}(x)|^2}{37} 
\eeqn
where $M_R$ is the mass of the resonance, 
$x \equiv 1-\frac{M_R^2}{M_{\jpsi}^2}$ and the function $H_{PS}(x)$ is
explicitly evaluated in ref. \cite[Eq. 17]{cakir94}. From Eq. \ref{ps} we obtain

\beqn \plabel{ls} BR (\eta(1760) \rightarrow gg)\; BR(\eta(1760) \rightarrow
\rho^0\rho^0) = \frac{6.6
\pm 4.6\; MeV}{60\pm 16\; MeV}
= 0.11 \pm 0.08
\eeqn
To deduce the two--gluon coupling of $\eta(1760)$, its $\rho^0\rho^0$ coupling needs to be
estimated.  We assume na\"{\i}ve branching ratios for $\eta(1760)\rightarrow \rho\rho,\;
\omega\omega$ which are consistent with the experimental value of
$BR(\jpsi\rightarrow\gamma\rho\rho)/BR(\jpsi\rightarrow\gamma\omega\omega)
= (4.5\pm 0.8) / (1.59\pm 0.33) \approx 3$, i.e. we take 
$BR(\eta(1760) \rightarrow \rho^0\rho^0) \lapprox \frac{1}{4}$.  From Eq. \ref{ls}
we hence deduce that $ BR (\eta(1760)
\rightarrow gg) \gapprox 0.4\pm 0.3$. Following ref. \cite{farrar96} the favoured conclusion is that
$\eta(1760)$ has gluonic admixture and is not pure $Q\bar{Q}$, even though the latter
possibility is allowed given the size of experimental errors. 
If the $\eta(1760)$ is $Q\bar{Q}$, it is expected to be the second radially excited (3S)
$\eta$. Such a state at 1.8 GeV decays dominantly
to $\rho\rho$ with a branching ratio of approximately $40\%$ 
(see Table B4 of ref. \protect\cite{page96rad}), 
yielding a $4\pi$ signal.
It should be noted that if the state is {\it pure} 3S $Q\bar{Q}$,   
Eq. \ref{ls} implies that $BR(\eta(1760)\rightarrow gg) \gapprox 0.8\pm 0.6$, which weakens the
hypothesis (here we used a 
branching ratio to $\rho^0\rho^0$ of $\frac{1}{3}\; 40$\%).
So we expect mixing with either the 
predicted glueball at $2.22\pm 0.32$ GeV \cite{teper96} or the hybrid
at $\approx 1.8 - 1.9$ GeV \cite{swanson95}. 
Although it is possible that there is significant
glueball mixing given the substantial errors in the mass estimate \cite{teper96}, 
the nearness of $\eta(1760)$ to
predicted hybrid masses, and the necessity of an isoscalar analogue to the hybrid candidate
$\pi(1800)$ (see below), henceforth motivate us to concentrate on  hybrid admixture in 
$\eta(1760)$.

It is also possible to obtain an upper bound \cite{cakir94} on $BR(\eta(1760)\rightarrow gg)$ if
we assume the CUSB limit \cite{pdg96,tuts} of $BR(\Upsilon\rightarrow \gamma X) \lapprox 8\times 10^{-5}$
for inclusive resonance production. The relevant relation, similar to Eq. \ref{ps}, is \cite{farrar96}

\beqn \plabel{up}
BR(\Upsilon\rightarrow \gamma R) = \frac{4\alpha}{5\alpha_S}\; (1-2.6\frac{\alpha_S}{\pi})\;
\frac{M_R}{M_{\Upsilon}^2}\; \Gamma(R\rightarrow gg)\; \frac{x|H_{PS}(x)|^2}{8\pi(\pi^2-9)} 
\eeqn
where $x \equiv 1-\frac{M_R^2}{M_{\Upsilon}^2}$. Taking $\alpha_S(m_b) = 0.18$,
we estimate that $BR(\eta(1760)\rightarrow gg)\lapprox 0.9\pm 0.3$, although we note that $H_{PS}(x)$ may
be unreliable for the value of $x$ used \cite{farrar96}. The upper bound is sensible and constrictive,
underlining the validity of the methods used, and indicating that the window for detection
of $\eta(1760)$ in $\Upsilon$ radiative decay at a B--factory could be small.

Even if the experimental information from Mark III and DM2 \cite{rho,omega} is unreliable, there is a
second indicator of the gluonic admixture in $\eta(1760)$.
Persuasive evidence \cite{ves} that the $\pi(1800)$ is dominantly an isovector hybrid
\cite{page96panic} has recently
emerged in diffractive $\pi$ production.
The state has width $212\pm 37$ MeV \cite{pdg96}, and detailed ratios of widths for its various decay modes
are known\footnote{The widths of $\pi(1800)$ are assumed to be in the ratio \protect\cite{ves,private}
1 ($f_0(980)_{\pi^+\pi^-}\pi^-$), $0.6\pm 0.2$ ($f_0(1300)_{\pi^+\pi^-}\pi^-$), $0.3\pm 0.1$ ($K^+K^-\pi^-$),
$0.4\pm 0.1$ ($a_0^-(980)_{\eta\pi^-}\eta$), $0.12\pm 0.05$ ($\eta\eta^{'}\pi^-$), 
$0.04\pm 0.02$ $f_0(1500)_{\eta\eta}\pi^-$, $< 0.18$ ($\rho^0\pi^-$), $< 0.06$ ($K^*K$) and
$0.4\pm 0.2$ ($\rho^-\omega$). If a mode is in subscript the branching ratio to 
the mode has not been included. Correcting for these branching ratios  \protect\cite{pdg96}
we estimate the width ratios
$1.9 \pm 0.1$ ($f_0(980)\pi$),  $0.9\pm 0.3$ ($f_0(1300)\pi$),  $1.0\pm 0.3$ ($K^*_0(1430)K$),  $0.9\pm 0.2$ ($a_0(980)\eta$),  $0.12\pm 0.05$ ($\eta\eta^{'}\pi$),  $ 0.6\pm 0.3$ ($f_0(1500)\pi$),  $< 0.36$ ($\rho\pi$),  
$<0.06$ ($K^*K$) and  $0.4\pm 0.2$ ($\rho\omega$).}.
The $\pi(1800)$ has strong hybrid features, including suppressed
decays to S--wave mesons \cite{page96panic,sel,page95light}, and 
is inconsistent with a 3S $Q\bar{Q}$ interpretation  \cite{page96rad}. Moreover,
an especially interesting feature of $\pi(1800)$ is its decay to $f_0(1500)\pi$ \cite{ves},
where $f_0(1500)$ is observed in the small $\eta\eta$ decay mode. We
estimate from experiment that $BR(\pi(1800)\rightarrow f_0(1500)\pi)
\approx 10\pm 6 \%$ (see Footnote 1), indicating significant decay to the glueball candidate $f_0(1500)$. 
A new analysis \cite{abele96} implies that the branching ratio can
be two times larger.
Amongst the dominant decay modes of $\pi(1800)$ 
is $f_0(980)\pi$, as well as below threshold $K^*_0 (1430) K$, yielding a
significant $KK\pi$ coupling \cite{ves}. We naturally expect an 
isoscalar analogue of $\pi(1800)$. 
For the isoscalar analogue, the dominant decays are expected to be
to $a_0(1450)\pi$ and $a_0(980)\pi$, yielding a large $\eta\pi\pi$ and 
a substantial $KK\pi$ width.
It is tantalizing that the
$\jpsi\rightarrow\gamma a_0(980)\pi$ data in ref. \cite{k1}
appear to have resonance behaviour in the $0^{-+}$ wave in the region $1.8 - 1.9$ GeV  
with width 0.1 GeV, most notably in Fig. 7a, and that the branching ratio to $a_0(980)\pi$
appears to be similar to that of $\eta(1440)$. Moreover, Fig. 1a of ref. \cite{k2} appears to be
 consistent with an excess of
$K^0_SK^{\pm}\pi^{\mp}$ events at $\sim 1.9$ GeV. It is hence surprising that refs. \cite{k1,k2}
make no reference to structures in the  $1.8 - 1.9$ GeV  region. The observations are consistent with
expectations for an isoscalar analogue of the $\pi(1800)$. 
The VES collaboration has put
a limit on the $K^* K$ signal for $\pi(1800)$ (see Footnote 1). It is imperative that the
behaviour of $K^* K$ in $\jpsi$ radiative decay in the 1.8 GeV region \cite[Fig. 7a]{k1} 
be studied in more detail to accertain whether there is also weak coupling to $K^* K$.

The $KK\pi$ data  \cite{k1,k2} can be explained as arising from the presence of 
the isoscalar analogue of $\pi(1800)$ as a component of $\eta(1760)$.
If this is indeed the case, and if the observations of $\eta(1760)$ 
in $\rho\rho$ and $\omega\omega$ \cite{rho,omega} are reliable,
we would also expect a $Q\bar{Q}$ component in $\eta(1760)$. This is because the isoscalar analogue 
of $\pi(1800)$ only couples weakly to $\rho\rho$ and $\omega\omega$, due
to the small coupling of $\pi(1800)$ to $\rho\omega$ (see Footnote 1).

We now turn to the theoretical estimation of $\Gamma(R\rightarrow gg)$, where $R$ is a 
hybrid.
The production of a hybrid is calculated assuming that one of the
gluons radiated in $\jpsi\rightarrow\gamma R$ acts as a
spectator. 

The physical picture for $\Gamma(R\rightarrow gg)$ is described as follows. For a hybrid meson which possesses an
excited gluonic degree of freedom, the quark and antiquark reside in
a colour--octet state instead of a colour singlet state as for regular mesons. 
For convenience, we evaluate $\Gamma(R\rightarrow
gg)$ in the valence--gluon picture \cite{leyaouanc85}. In this description, the colour state of
the hybrid can be written as $|\bar Q_i\lambda^a_{ij}Q_jA^a>$ where we suppress other quantum
numbers such as spin.

The interaction for $R(\bar QQg)\rightarrow gg$ can be depicted as an annihilation
of $Q\bar Q$ into a gluon with the valence gluon acting as a spectator. This
picture is similar to semileptonic and non--leptonic decays of mesons, which
are discussed in ref. \cite{Bauer}. In meson decays, a quark 
(antiquark) undergoes a weak transition, while another antiquark (quark)
maintains its identity unchanged as a spectator, although it was in the parent 
meson and later transits to the daughter meson.  Although its kinematic states in
the parent and daughter mesons are very different, it is believed that a 
non--perturbative QCD effect of order unity can make them match. 
In our case, the valence gluon is a spectator while the $Q\bar Q$ turn
into an off--shell gluon. The quark diagram is shown in Figure 1.
Both the gluon coming from the $Q\bar{Q}$ and the spectator gluon 
are off--shell. The two off--shell gluons can exchange
many soft gluons via non--perturbative QCD interactions and finally turn into
two free gluons emerging as the final colour--singlet state. This non--perturbative
effect provides a factor whose precise evaluation is beyond our
present ability, but as argued elsewhere \cite{Bauer}, must be of order 
unity. 
We take the factor $O(1)$ to be 1 in our numerical calculations.

\begin{figure}
\begin{center}
\leavevmode
\hbox{\epsfxsize=4 in}
\epsfbox{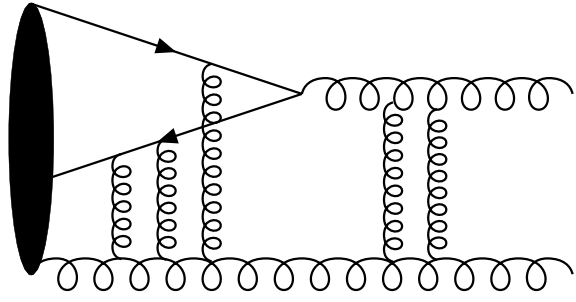}
\caption{Non--perturbative interactions of $O(1)$ in the spectator model.}
\end{center}
\end{figure}

Now we need to evaluate the subprocess pertaining to the annihilation of $(Q\bar Q)^a
\rightarrow g^a$. It is well known that $e^+e^-$ annihilation into a
final state $F$ can be realized via an off--shell photon as
\begin{equation}
\sigma(e^+e^-\rightarrow F)={4\pi\alpha\over s^{3/2}}\Gamma(\gamma^*
\rightarrow F)
\end{equation}
where $\sqrt{s}$ is the s--channel invariant mass. For example, for $e^+e^-\rightarrow\mu^+\mu^-$, one has
\begin{equation}
\sigma(e^+e^-\rightarrow\mu^+\mu^-)={4\pi\alpha^2\over 3s}
\end{equation}
where 
\begin{equation}
\Gamma(\gamma^*\rightarrow\mu^+\mu^-)={\sqrt s\over 3}\alpha
\end{equation}
is the decay width of an off--shell virtual photon into a $\mu^+\mu^-$ pair.
The expression neglects the muon mass.  

Since $Q\bar Q$ reside in a bound state and constitute a 
colour octet, a form factor is needed to reflect the binding effect. In
a non--relativistic scenario \cite{book}, one can describe this as
\begin{eqnarray}
<0|J|Q\bar Q> &\propto & \int d^3p_1d^3p_2\psi(\vec p_1-{1\over 2}\vec p_c,
\vec p_2-{1\over 2}\vec p_c)\delta(\vec p_1+\vec p_2-\vec p_c)
\nonumber\\
&=& (2\pi)^3\psi(\vec r_1=0,\vec r_2=0)
\end{eqnarray}
which is just the wave function at the origin. Here $\vec{p}_1$ is the 3--momentum of the quark, and $\vec{p}_c$ the
centre of mass momentum of the $Q\bar{Q}$ system.

In a close analogy, considering colour and other quantum numbers
with a proper normalization, we obtain
\begin{equation}
\Gamma(0^{-+}\; Q\bar Qg\rightarrow gg) 
= {16\pi(s+2m^2)\alpha_S\over s^2(s-4m^2)^{1/2}}\;\Gamma_{g} |\psi_H(0)|^2
\end{equation}
where
\begin{eqnarray}
\psi_H(0) = 
{1\over R^{3/2}} 
\end{eqnarray}
and $R$ is the radius associated with the simple harmonic oscillator (S.H.O.) wave function.

If $\Gamma_g$ is the decay width of the off--shell gluon, we have similarly to
$\gamma^*\rightarrow \mu^+\mu^-$ that
\begin{equation}
\Gamma_g={4(M_{Q\bar{Q}}^2-4m^2)^{1/2}\alpha_S\over 3 M_{Q\bar{Q}}^2}\; 
(M_{Q\bar{Q}}^2+2m^2),
\end{equation}
where the extra factor of 4 is due to quark colour and $M_{Q\bar{Q}}$ is the mass
of the $Q\bar{Q}$ system. 

If we assume that $\sqrt s =M_{Q\bar{Q}}$, then
$\sqrt s = M_{Q\bar{Q}} = 1 - 1.5$  GeV.
Thus we have
\begin{equation}
\Gamma(0^{-+}\; Q\bar Qg\rightarrow gg)={16\pi(s+2m^2)\alpha_S\over s^2(s-4m^2)^{1/2} R^3}
\; \Gamma_g
\end{equation}
where
$m$ is the constituent
quark mass of about 0.33 GeV. Note that $BR(0^{-+}\; Q\bar Qg\rightarrow gg)
\sim \Gamma(0^{-+}\; Q\bar Qg\rightarrow gg)/\Gamma_g \sim O(\alpha_S)$ as expected for a hybrid.
For $\alpha_S = 0.3-0.4$ and $R^2 = 10 - 12\;$ GeV$^{-2}$ \cite{leyaouanc85}
we find
$\Gamma(0^{-+}\; Q\bar Qg\rightarrow gg)\sim 180\pm 110$ MeV.

It should be noted that all the expressions given above are based on relativistic
field theory, except for the wavefunction at the origin which stands for the 
necessary
form factor. This picture is almost universally adopted \cite{farrar96}. 

For conventional meson production in the non--relativistic limit \cite{farrar96,cakir94,book}
\begin{equation}
\Gamma(0^{-+}\; 3S\; Q\bar{Q}\rightarrow gg)={8\over 3}{\alpha_S^2\over M_{R}^2}
|\Psi_{3S}(0)|^2 = 5{\alpha_S^2\over M_{R}^2}
{1\over R^3}
\end{equation}
where for the purpose of calculation we use the na\"{\i}ve S.H.O. expression
\begin{equation}
\Psi_{3S}(0)=\sqrt{\frac{15}{8R^3}}
\end{equation}
We obtain for the same parameters as before that
${\Gamma(0^{-+}\; 3S\;  Q\bar{Q}\rightarrow gg)} \sim 6\pm 2$ MeV. 
In addition, if we assume that the $gg$ coupling of 3S is similar to 2S,
we can obtain from $BR(\eta(1295)\rightarrow gg) \sim 0.25$ \cite{farrar96}
that $\Gamma(0^{-+}\; 3S\;  Q\bar{Q}\rightarrow gg)\sim 10$ MeV.

We deduce that the two--gluon coupling of a hybrid is considerably larger
than that of 3S $Q\bar{Q}$. This implies from Eqs. \ref{ps} and \ref{up} that the branching ratio of a
hybrid in radiative $\jpsi$ or $\Upsilon$ decay is substantial relative to that of 3S $Q\bar{Q}$.

We shall now compare two--gluon to two--photon coupling. 
For isoscalar mesons, written in flavour singlet 
and octet
components as $\cos\theta|8\rangle+\sin\theta|1\rangle$, the ratio between the  
$\gamma\gamma$ and $gg$ coupling is \cite{farrar96,kwong}

\beqn
\frac{\Gamma(R\rightarrow\gamma\gamma)}{\Gamma(R\rightarrow gg)}
= \frac{1}{4}(\frac{\alpha}{\alpha_S})^2 \frac{1}{1+\frac{8.2\alpha_S}{\pi}}
\frac{\cos^2(\theta-\tau)}{\cos^2\theta}
\eeqn
where $\tau\equiv\tan^{-1}\frac{1}{2\sqrt{2}}\approx 19.5^o$. 
Since this relation is only valid for $Q\bar{Q}$, and since we already
have a lower bound for $\Gamma(\eta(1760)\rightarrow gg)$, we can use it to 
obtain a lower bound on $\Gamma(\eta(1760)\; Q\bar{Q}\rightarrow\gamma\gamma)$. Thus if
experiment finds a $\gamma\gamma$ coupling below this value, the state is
not pure $Q\bar{Q}$, since $Q\bar{Q}$ are expected 
to be strongly coupled to $\gamma\gamma$ \cite{page96gamma}. Using the value $\alpha_S\approx 0.48\pm 0.05$
obtained by fitting the above relation to the data on $f_2(1270)$ and $f_2(1525)$
\cite{farrar96}, together with the value $\Gamma(R\rightarrow gg)\gapprox 4\times (6.6\pm4.6)$ MeV
(see Eq. \ref{ls}), we find 

\beqn
\Gamma(\eta(1760)\; Q\bar{Q}\rightarrow\gamma\gamma)\gapprox (0.68\pm 0.50)
\frac{\cos^2(\theta-\beta_c)}{\cos^2\theta} \;\; \mbox{keV}
\eeqn 
For a state containing no strange quarks ($\theta=35.3^o$) we obtain
$\Gamma(\eta(1760)\; Q\bar{Q}\rightarrow\gamma\gamma)\gapprox (0.9\pm 0.7)$ keV. 
For the partner $s\bar{s}$ state ($\theta=-54.7^o$) production is
suppressed: 
$\Gamma(\eta(1760)\; Q\bar{Q}\rightarrow\gamma\gamma)\gapprox (0.15\pm 0.11)$ keV.
$Q\bar{Q}$ two--photon widths of $O($keV) would generate a 
prominent signal in $\gamma\gamma\rightarrow 4\pi$. Specifically, if the
branching ratio to $\rho\rho$ is $40\%$ as mentioned before,  
we have $\Gamma(\gamma\gamma\rightarrow\eta(1760)\rightarrow 4\pi) \gapprox 0.4\pm 0.3$ keV.
Conversely, the confirmation of $BR(\jpsi\rightarrow\gamma\eta(1760))
 \gapprox 4 \times (0.13\times 10^{-3})$ together
with the absence of a signal in $\gamma\gamma$ at $\gapprox 0.9$ keV 
could support the $\eta(1760)$ as dominantly gluonic rather than $Q\bar{Q}$.

We now explicitly evaluate $\gamma\gamma$ widths. 
From information on the $\rho\omega$ coupling of $\pi(1800)$ we
can make an estimate of $\gamma\gamma$ coupling of $\pi(1800)$ and $\eta(1760)$ using 
VMD, independent on whether the state is a hybrid or $Q\bar{Q}$. We use the following 
VMD relation 
pertaining to states containing only $u$ and $d$ quarks,
which incorporates the effects of phase--space

\beqn
\plabel{gam}
\Gamma(R\rightarrow\gamma\gamma) = (\frac{\pi\alpha}{\gamma^2_{\rho}})^2 \; 
\frac{\Gamma{(\pi(1800)\rightarrow\rho\omega)}}{8(1-4(\frac{m_{\rho\omega}}{M_R})^2)^{L+\frac{1}{2}}}
 \; (\frac{M_R}{\tilde{M}_B})^2  
\;\times\; \left\{ \barr{cl} 4 {\cal R}^2_{\omega} & \mbox{for } \frac{1}{\sqrt{2}}(u\bar{u}-d\bar{d}) \\ 
(1+{\cal R}^2_{\omega})^2 & \mbox{for } \frac{1}{\sqrt{2}}(u\bar{u}+d\bar{d})  \earr \right.
\eeqn
where we make the approximation that the mass of $\rho$ and $\omega$ are the same (with
$m_{\rho\omega}$ their average mass), and that the isoscalar and isovector
resonances have the same mass. ${\tilde{M}_B} = 720$ MeV takes account of hadronic
P--wave ($L=1$) phase space \cite{page95light,page96gamma}, ${\cal R}_{\omega} = 0.30$ \cite{page96gamma} denotes the photon
coupling of $\omega$ relative to that of $\rho$, and the VDM coupling is $\gamma^2_{\rho}/4\pi=0.507$ \cite{page96gamma}.
We can estimate that $\Gamma(\pi(1800)\rightarrow\rho\omega) = BR(\pi(1800)\rightarrow\rho\omega)\;
\Gamma_T(\pi(1800)) = 15\pm 8$ MeV (see Footnote 1). From Eq. \ref{gam} we obtain $\Gamma(\pi(1800)\rightarrow\gamma\gamma)
= 0.4\pm 0.2$ keV and $\Gamma(\eta(1760)\rightarrow\gamma\gamma) = 1.4\pm 0.7$ keV. The predicted
$\gamma\gamma$ coupling of $\eta(1760)$ is sizable, 
making its detection in the $KK\pi$ and $\eta\pi\pi$ channels a 
realistic prospect. If these expectations are correct, comparison with 
$\Gamma(\eta(1760)\; Q\bar{Q}\rightarrow\gamma\gamma)\gapprox 0.9\pm 0.7$ keV
derived above leaves the issue of the gluonic content of $\eta(1760)$ unresolved. Improvement
in errors should have significant consequences here.


It has been argued that $\pi(1800)$ can influence $p\bar{p}$ annihilation data \cite{private}
and enhance CP violating modes in Cabibbo suppressed $D^0$ decays \cite{lipkin95} due to
mixing with $p\bar{p}$ and $D$ which both have similar mass. 
Clearly the $\eta(1760)$ will have similar effects.
Especially interesting here are the $\eta\pi\pi$ and $4\pi$ channels which cannot result from mixing
with $\pi(1800)$. Also interesting is that $D^0\rightarrow (K^0K^-\pi^+)_S$ and $ (\bar{K}^0K^+\pi^-)_S$
are superficially different \cite{pdg96}, though with large errors, possibly indicating CP violation.

The experimental data on $0^{-+}$ production in radiative $\jpsi$ decays need clarification and confirmation before strong conclusions can be drawn. This should be a high priority at a $\tau$--charm factory or BEPC.
An immediate challenge for experimental analysis includes confirmation of a resonance in the 1.8 GeV region
in $\jpsi\rightarrow \gamma\;  (KK\pi, \rho\rho, \omega\omega)$.
Comparison with signals in $\gamma\gamma\rightarrow 4\pi,KK\pi,\eta\pi\pi$ should
also be made at facilities like Babar, CLEO II, LEP2 and LHC.

Progress towards the resolution of the nature of $\eta(1760)$ can be made by using the
techniques presented.

\vspace{.6cm}

Helpful discussions with D.V. Bugg, F.E. Close, J. McGovern, P. Sutton, A.M. Zaitsev and B.S. Zou are acknowledged.
XQL thanks the Theory Group at RAL for the provision of a stimulating working environment.

\appendix


%

\end{document}